\documentclass{iaus}
\usepackage{graphicx}

\title[IAU 249: Comparison of cloud models ] 
{Comparison of cloud models\\ for Brown Dwarfs}

\author[Christiane Helling]   
{Ch.~Helling$^{1}$, A.~Ackerman$^{2}$,
 F.~Allard$^{3}$, M.~Dehn$^{4}$, P.~Hauschildt$^{4}$,
D.~Homeier$^{5}$, K.~Lodders$^{6}$, M.~Marley$^{7}$,
F.~Rietmeijer$^{8}$,  T.~Tsuji$^{9}$, P.~Woitke$^{10}$}

\affiliation{
$^1$SUPA, School of Physics \& Astronomy, University of St Andrews, North Haugh, St Andrews,
  KY16 9SS, Scotland, UK\\ email: {\tt Christiane.Helling@st-andrews.ac.uk}\\[\affilskip]
$^2$ NASA Goddard Institute of Space Studies, New York, USA\\[\affilskip]
            $^3$ Centre de Recherche Astrophysique de Lyon, Universit\'e de  Lyon, France \\[\affilskip]
            $^4$ Hamburger Sternwarte, Hamburg, Germany\\[\affilskip]
            $^5$ Universit\"at G\"ottingen, Institut f\"ur Astrophysik, G\"ottingen, Germany \\[\affilskip]
            $^6$ Planetary Chemsitry Laboratory, Department of Earth and Planetary Sciences,\\ Washington University, St. Louis, USA\\[\affilskip]
            $^7$~NASA Ames Research Center, Moffett Field, USA \\[\affilskip]
            $^8$ Department of Earth and Planetary  Sciences, University New Mexico, Albuquerque, USA \\[\affilskip]
            $^9$ Institute of Astronomy, The University of Tokyo, Mitaka, Tokyo, Japan\\[\affilskip]
            $^{10}$ UK Astronomy Technology Centre, Royal Observatory,  Edinburgh, UK
}
				   
\pubyear{2008}
\volume{xxx}  
\pagerange{119--126}
       \jname{{\sf Exoplanets:} Detection, Formation, Dynamics}
			     \editors{??}

\begin{document}

\maketitle

\begin{abstract}
A test case comparison is presented for different dust cloud model
approaches applied in brown dwarfs and giant gas planets.  We aim to
achieve more transparency in evaluating the uncertainty inherent to
theoretical modelling. We show in how far model results for
characteristic dust quantities vary due to different assumptions. We
also demonstrate differences in the spectral energy distributions
resulting from our individual cloud modelling in 1D substellar
atmosphere simulations.  
\end{abstract}

\firstsection 

\begin{figure}[h]
\begin{center}
\resizebox{0.75\textwidth}{!}{\includegraphics{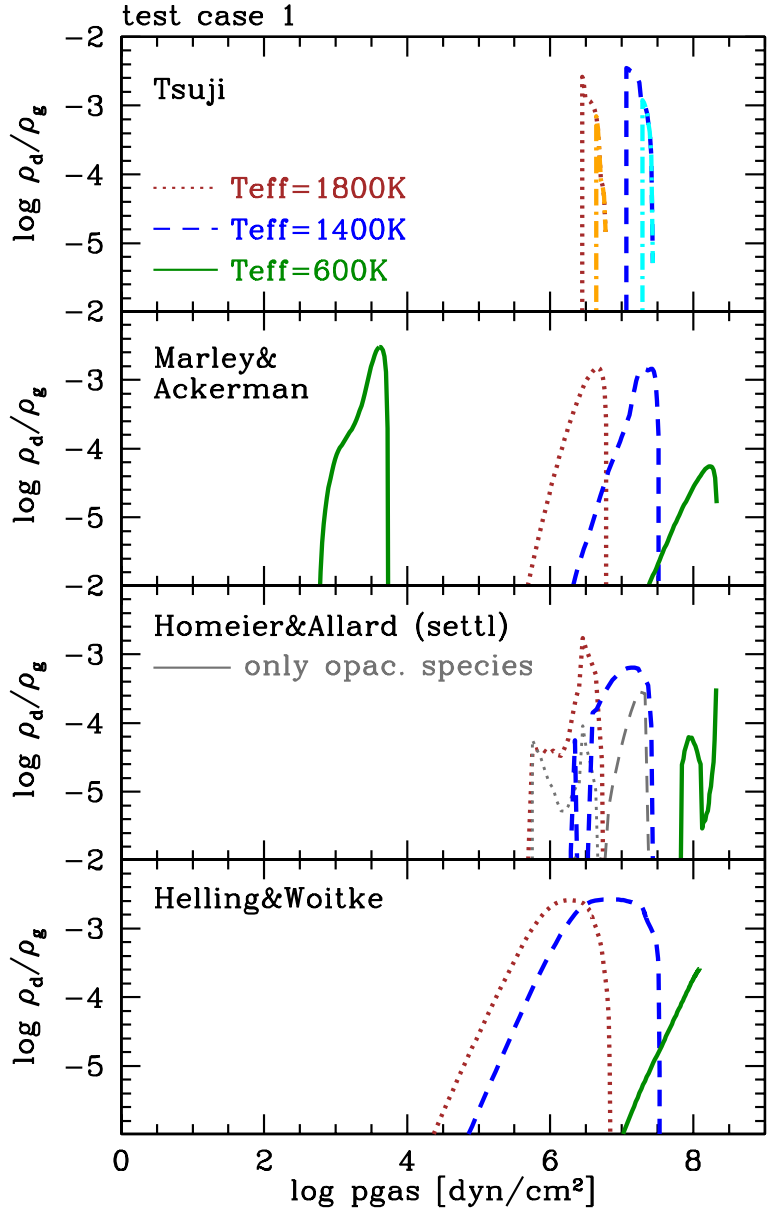}}
 \caption{Results for {\bf Test case 1} (L and
T-dwarf regime): The dust content (dust-to-gas ratio $\rho_{\rm
d}/\rho_{\rm g}$)
of the clouds calculated by different model
approaches. Note that in the Tsuji-case  two models are
plotted for each T$_{\rm eff}$, T$_{\rm crit}$=1700K (red/blue) and
T$_{\rm crit}$=1900K (orange/cyan).}
   \label{fig1}
\end{center}
\end{figure}

\section{Introduction}
Spectra are windows through which we access the physics and the
chemistry of substellar atmospheres down into the planetary
regime. Consequently, model atmospheres are needed to produce
synthetic spectra. The complexity of such simulations necessitates the
use of approximations and/or analytic simplifications to make them
numerically feasible. The input data needed can be frequently
incomplete or of limited accuracy, translating into uncertainties in
the model results.  The formation of clouds in their atmospheres is
one of the most fundamental challenges in studies of planets and brown
dwarfs.  Element depletion and gravitational settling are only two
feedback mechanisms arising.  We therefore have set out to compare our
approaches of cloud formation in substellar atmosphere simulations to
achieve more transparency in evaluating the uncertainty inherent to
theoretical modelling. We show in how far model results for
characteristic dust quantities vary due to different
assumptions. We also demonstrate differences in the spectral energy
distributions resulting from our individual cloud modelling in 1D
substellar atmosphere simulations.

\section{Cloud models \& Test cases}\label{s:models}

Theoretical models of substellar atmosphere aim to describe the
existence of clouds made of small particles and their influence on the
atmosphere's energy budget. Basically two model philosophies are
followed: The cloud particles hoover inside the atmosphere (Tsuji
2005) or, the particles gravitationally settle, hence disappear from
the atmosphere but need to re-form (Ackerman \& Marley 2001; Sudarsky
et al. 2003, Woitke \& Helling 2004, Allard et al. (2007). All models
aim to produce as accurate as possible synthetic spectra by modelling
cloud characteristics like the number of dust particles, the material
composition of the cloud grains/droplets, material composition, and
the size of the individual cloud particles.  In general, formation of
clouds by condensation would be controlled by kinetic factors, for
example induced by turbulence, and amorphous solids would form. Given
sufficient time and energy, these amorphous solids will crystallise
and form equilibrium solids known as minerals. Phase-equilibrium is
adopted in the approaches of Marley \& Ackerman (2001), Tsuji (2005),
Allard et al. (2007), and the kinetic approach is taken by Woitke \&
Helling (2004) and Helling et al. (2007).

\noindent
The dust cloud models involved in the test calculations are summarised
(and simplified) as follows:

\smallskip
\begin{tabular}{lp{0.3cm}l}
{\bf Fixed grain size} &&  {\bf Time scale comparison}\\
(Tsuji 2005)&&
(Allard et al. 2007)\\
-- grain size $a$  $=$ const &&  $\tau_{\rm mix}=\tau_{\rm sed}(a)\,\Rightarrow\, (T,p, a)$\\
-- between T$_{\rm cond}$ and T$_{\rm cr}$ &&  $\tau_{\rm gr}(T,p,a) < \tau_{\rm mix}\,\Rightarrow\, \tau_{\rm mix}=\tau_{\rm gr}(a)$\\
&&  $\tau_{\rm gr}(T,p,a) > \tau_{\rm mix}\,\Rightarrow\,\tau_{\rm mix}=\tau_{\rm cond}(a)$\\
\hline
{\bf Diffusive transport} && {\bf Top - Down approach}\\
(Marley \& Ackerman 2001;  && (Woitke \& Helling 2004,\\
Marley et al. 2007) && Helling, Woitke, Thi 2007)\\
--  diffusive transport of homog. dust && -- conserv. equation for dirty dust format.\\ 
-- needs dust formation parameters.   && + convective over-shooting\\
\end{tabular}

\medskip
\noindent
We designed test cases to study difference arising from our different
theoretical approach to model clouds in substellar atmosphere
simulations. We disentangle our individual cloud models from the
complete atmosphere problem ({\bf Test case 1}) in order to exclude possible
feedback amplifications in the entire model atmosphere codes.  Only as
a second step, we investigate possible differences in the spectral
appearance of a given substellar object ({\bf Test case 2}). The test cases
are designed as follows and more details can be found under
{http://phoenix.hs.uni-hamburg.de/BrownDwarfsToPlanets1/}:

\smallskip
\begin{tabbing}
{\bf Test case 1:} \hspace*{4.5cm} {\bf Test case 2:}\\
 -- \,\,\,\=  compare results   from    \hspace*{3cm} \= -- \,\,\,\= compare results  from \\
    \> cloud models only                         \>   \>  complete 1D atmosphere simulation \\
 --  \>{\it local} quantities given :                  \> -- \>  {\it global} quantities given:\\
     \> (T, p ,$v_{\rm conv}$) structure         \>  \>T$_{\rm eff}=1800$K, $\log\,g=5.$\\
 --  \>different dust cloud treatment            \> --  \> different model atmosphere codes\\
     \> \> \> including different dust cloud treatment
\end{tabbing}

\begin{figure}[t]
\begin{center}
\resizebox{0.67\textwidth}{!}{\includegraphics{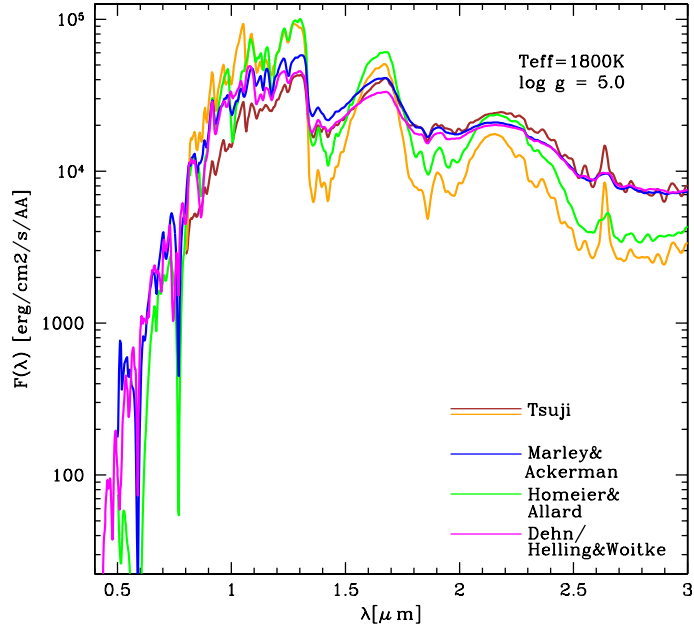}}
 \caption{Results for {\bf Test case 2} (L dwarf): Spectral energy
distribution between 0.5--3.0$\mu$m from different atmosphere codes
utilising different cloud-model approaches for (T$_{\rm eff}=1800$K,
$\log\,g=5.0$, [M/H]=0.0). Note, that the two Tsuji-cases
demonstrate the influence of the cloud thickness: T$_{\rm crit}$=1700K
(brown) -- thick cloud, T$_{\rm crit}$=1900K (orange) -- thin cloud.}
   \label{fig2}
\end{center}
\end{figure}

\section{Results}
 We perform our tests for brown dwarfs of spectral class L and T.
Figure~\ref{fig1} shows the dust content (dust-to-gas ratio $\rho_{\rm
d}/\rho_{\rm g}$) of the clouds calculated by different model
approaches (Sect.~\ref{s:models}). We observe similarities amongst the
different cloud model results like the location of the inner cloud
edge and the approximate location of the maximum dust content inside
the cloud. However, differences are apparent like e.g. the pressure
range covered by the clouds. We find the same behaviour for other
opacity relevant cloud quantities like grain size and cloud material
composition. It is therefore not surprising that the resulting spectra
(Fig.~\ref{fig2}) show a very similar general behaviour, like the
absolute flux level, but vary in details. Figure~\ref{fig2} also
demonstrates that the model atmosphere simulations for T$_{\rm
eff}=1800$K, $\log\,g=5.0$, [M/H]=0.0 fall into two groups: those
producing a thin cloud layer (Tsuji with T$_{\rm crit}$=1900K,
Homeier\,\& Allard) and those producing a thick cloud layer (Tsuji with
T$_{\rm crit}$=1700K, Marley \& Ackerman, Dehn/Helling\&Woitke (Dehn 2007)). It is
interesting to realise that two entirely different mechanisms are
responsible to move from one group to the other: Tsuji varies the
geometrical cloud thickness by using a critical temperature T$_{\rm
crit}$ for a constant grain size which results in an optically
thinner/thicker cloud. The cloud models employed by Homeier\,\& Allard,
Marley \& Ackerman, Dehn/Helling\&Woitke produce  different amounts of dust
(Fig.~\ref{fig1}) and different grain size distributions across the
cloud height causing an optically thiner/thicker cloud (not shown).

\section{Conclusion}
Our test case studies show that the results of our individual cloud
models are comparable regarding general feature like the location of
cloud base and the maximum dust-to-gas ratio. However, the cloud-model
results differ if studied in more detail. It is therefore no surprise
that the spectral energy distributions produced from our different 1D
atmosphere simulations for a given parameter combination (T$_{\rm
eff}$, $\log$g, [M/H]) differ in almost the entire wavelength
range. It remains to quantify these differences in order to provide a
general range of applicability for e.g. stellar parameter determinations
inside and beyond the substellar regime.

\medskip
\noindent
{\small\bf Acknowledgement:}\\ We thank all
the participants of the workshop {\sf From Brown Dwarfs to Planets:
Chemistry and Cloud formation} which was supported by the Lorentz
Center of the University Leiden, Nederlandse Organisatie voor
Wetenschappelijk Onderzoek, The Netherlands research School for
Astronomy, the Scottish University Physics Alliance, and European
Space Agency. ChH acknowledges an IAU travel grant.

\end{document}